\pdfoutput=1 % only if pdf/png/jpg images are used
\documentclass{JINST}

\usepackage{amsmath, amsthm, amssymb, float}
\DeclareMathAlphabet{\mathpzc}{OT1}{pzc}{m}{it}

\usepackage{fancyhdr}
\title{Liquid Argon TPC Signal Formation, Signal Processing and Hit Reconstruction}

\author{B. Baller$^a$\\
\llap{$^a$}Fermi National Accelerator Laboratory,\\
  Batavia, IL, USA\\
E-mail: \email{baller@fnal.gov}}

\abstract{This document describes the early stage of the reconstruction chain that was developed for the ArgoNeuT and MicroBooNE experiments at Fermilab. These experiments study accelerator neutrino interactions that occur in a Liquid Argon Time Projection Chamber. Reconstructing the properties of particles produced in these interactions requires knowledge of the micro-physics processes that affect the creation and transport of ionization electrons to the readout system. A wire signal deconvolution technique was developed to convert wire signals to a standard form for hit reconstruction, to remove artifacts in the electronics chain and to remove coherent noise. 
}

\keywords{liquid argon; TPC}

\begin{document}

\section{Introduction}\label{sec:intro}

The liquid argon time projection chamber (LAr TPC) is ideal for the study of neutrino interactions as it provides mm-scale resolution in a $\mathcal{O}$(100 m$^3$) scale volume. A LAr TPC also produces a prodigious amount of information that must be processed to extract $\mathcal{O}$(100) physics quantities in a neutrino interaction. As an example, the MicroBooNE detector \cite{ubdetector} has a volume of 62 m$^3$. It is instrumented to measure the charge deposited in voxels of size $\approx$10 mm$^3$ resulting in $6 \times 10^9$ measurements for each beam spill. Images of the data have a photographic quality that has been likened to that of a bubble chamber. This report focuses on the challenge of extracting physics objects from this large data sample.

Significant progress has been made in developing the technology as exemplified by physics results from ICARUS \cite{icarusdetector} and ArgoNeuT \cite{argodetector} and in the successful operation of test stands such as LAPD \cite{lapd}, LongBo \cite{longbo}, CAPTAIN \cite{captain}, ArgonTube \cite{argontube}, etc. Results were obtained from semi-automatic reconstruction of complex event topologies or fully-automatic reconstruction of simple events. The large data volumes generated by the current generation of  experiments such as MicroBooNE and LArIAT \cite{lariat}, and the need to reconstruct more complicated events motivates improving reconstruction performance.

Reconstructing isolated particles is routinely done with efficiencies approaching 100\%. Minimum ionizing particles deposit $\gtrsim$ 500 keV in a wire hit. Efficient reconstruction of tracks $\gtrsim$1 cm (proton kinetic energy $\gtrsim$ 20 MeV) is feasible using techniques that are described here. The difficulty arises in reconstructing tracks that are embedded within a cluster of charge deposited by other tracks. Achieving this capability enables exploring the role of final state interactions in neutrino - argon interactions. Reconstructing tracks in a high-density shower or in the vicinity of a deep inelastic neutrino interaction is a significant challenge.

We begin with a review of a subset of the physics processes that ensue when an ionization event occurs within the detector. 
Discussion in this section is limited to the processes of creation, loss and transport of ionization electrons that form signals on the TPC wires. Only passing mention is made of the use of a light collection system to provide a trigger. Online processing of wire signals by the readout electronics is discussed only insofar as it motivates the need for offline signal processing.  The reader is referred to Appendix \ref{sec:calc} that describes a  LAr TPC calculator that allows a more precise estimate of the order of magnitude estimates given in this section. 

Section \ref{sec:osp} describes the deconvolution method that corrects for the  response of the readout electronics. This technique was originally developed to overcome an unavoidable artifact of the ArgoNeuT electronics. This convolution method also provides a mechanism for correcting the seemingly complex direct and indirect currents induced on TPC wires by ionization electrons. This processing stage produces a standardized set of wire signals that have a Gaussian-like shape.

A hit finding algorithm that uses a Gaussian fit is described in Section \ref{sec:hit}. Here we make the observation that an unambiguous connection between a hit and a discrete ionization event cannot be made at this early stage of reconstruction.

The methods described in this report were developed over the last 8 years and are currently implemented in the LArSoft software suite. All data shown are real unless labeled ``simulated''.

\section{Signal Formation}\label{sec:tpc}

\begin{figure}[tbp] 
\centering
\includegraphics[width=.49\textwidth]{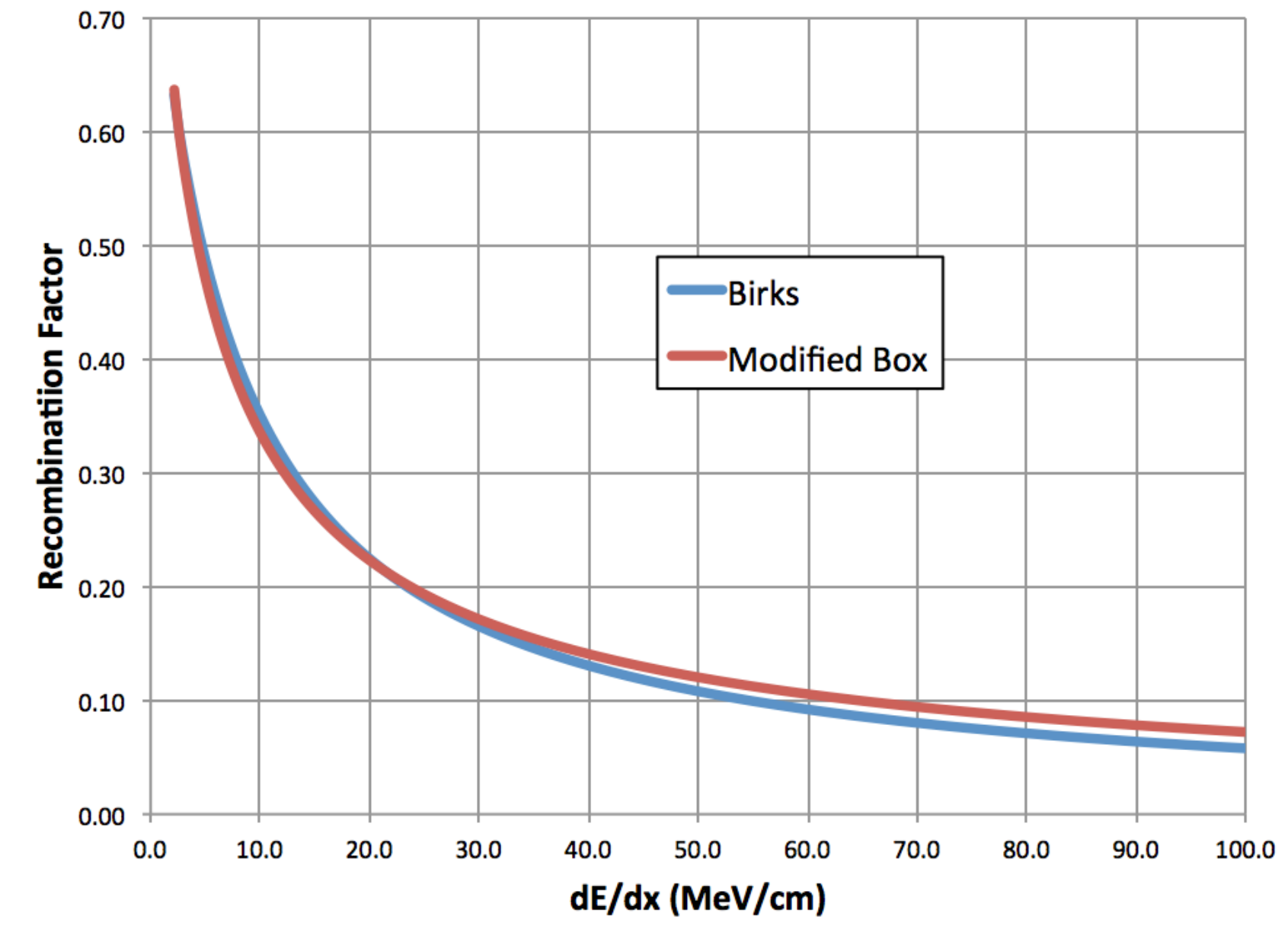}
\includegraphics[width=.49\textwidth]{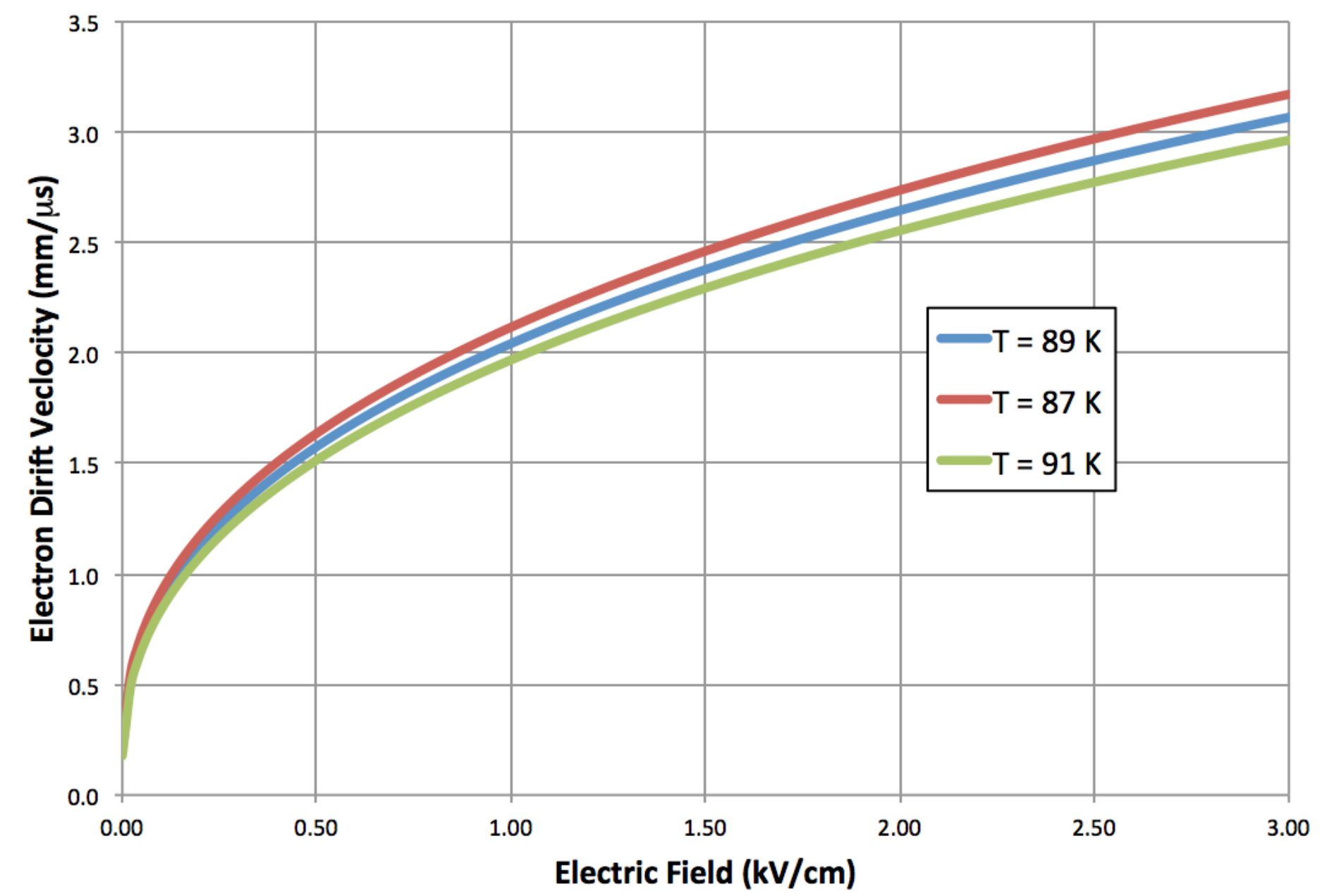}
\caption{Left: Birks and Modified Box Model parameterizations of electron recombination for $\mathcal{E}$ = 0.5 kV/cm. Right: Parameterization of the drift velocity of electrons in LAr as a function of electric field and temperature from reference \cite{driftvelocity}. }
\label{fig:driftvelocity}
\end{figure}

Electrons liberated by the passage of a particle in a LAr TPC are separated from their parent argon ions by $\approx2 \mu m$ after reaching thermal energies \cite{wojcik}. The electron and ion columns separate under the influence of the TPC electric field, $\mathcal{E}$, that is typically < 1 kV/cm. Electron-ion recombination occurs for the next few nanoseconds until the columns are well separated. The fraction of electrons that escape recombination, $\mathcal{R}$, can be modeled by a Birks ``law'' \cite{birk} \cite{icarrecomb} or alternatively by the ``Modified Box Model'' \cite{thomas} \cite{ballerrecomb} shown in equations \ref{birks} and \ref{box}. Both of these empirical models are based on the columnar theory of recombination by Jaffe \cite{jaffe} and provide similar performance. The $A_B$, $k_B$, $\alpha$ and $\beta$ parameters are found by fitting the model to the charge collected from stopping particles. The dependence on $dE/dx$ is shown in Figure \ref{fig:driftvelocity}. The $\mathcal{R}$ values for these two models differ by less than 10\% for $dE/dx$ < 35 MeV/cm and approaches 25\% at 100 MeV/cm.

\begin{equation}
\label{birks}
\mathcal{R}_{\rm{ICARUS}}  = \frac{A_B}{1 + k_B\cdot (dE/dx)/\mathcal{E}}
\end{equation}

\begin{equation}
\label{box}
{\mathcal R_{\rm{Box}}} = \frac{1}{\xi}ln(\alpha + \xi), \quad  \mathrm{where} \quad \xi = \beta (dE/dx) / \mathcal E.
\end{equation}

The separation time of the electron and ion columns also depends on the angle of the columns relative to the  $\mathcal{E}$ field direction. The two columns for an idealized particle traveling exactly in the $\mathcal{E}$ field direction, $\phi$ = 0, would overlap completely for many milliseconds resulting in complete recombination, $\mathcal{R} \approx 0$. The expected sin($\phi$) dependence is not included in the above equations because no significant angle dependence has been observed\cite{ballerrecomb}. A likely cause of this discrepancy is that the simple geometric model doesn't account for ionization from $\delta$-rays. 

%The columnar model predicts a significant increase in recombination when the particle travels close to the electric field direction with a sin($\phi$) dependence. This can be understood using a situation where an idealized particle travels exactly in the $\mathcal{E}$ field direction, $\phi$ = 0, without producing $\delta$-rays. The columns would overlap completely for all time resulting in complete recombination of all electrons and ions. Measurements of $\mathcal{R}$ using a sample of stopping protons \cite{ballerrecomb} shows a reduction of 5\% at $\phi = 40^\circ$ instead of the expected 25\% decrease.

\begin{figure}[H] 
\centering
\includegraphics[width=.5\textwidth]{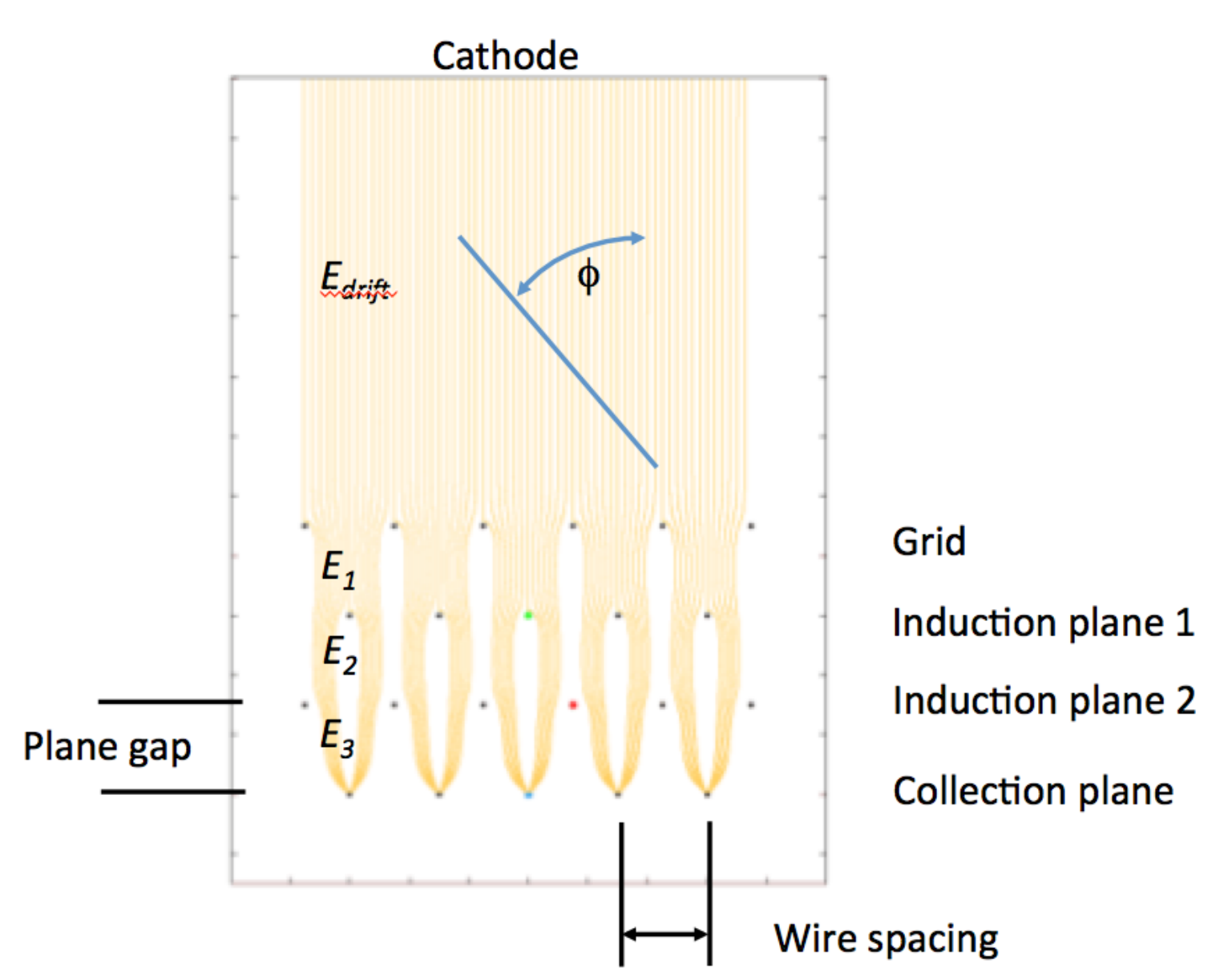}
%$\vcenter{\hbox{
\includegraphics[width=.4\textwidth]{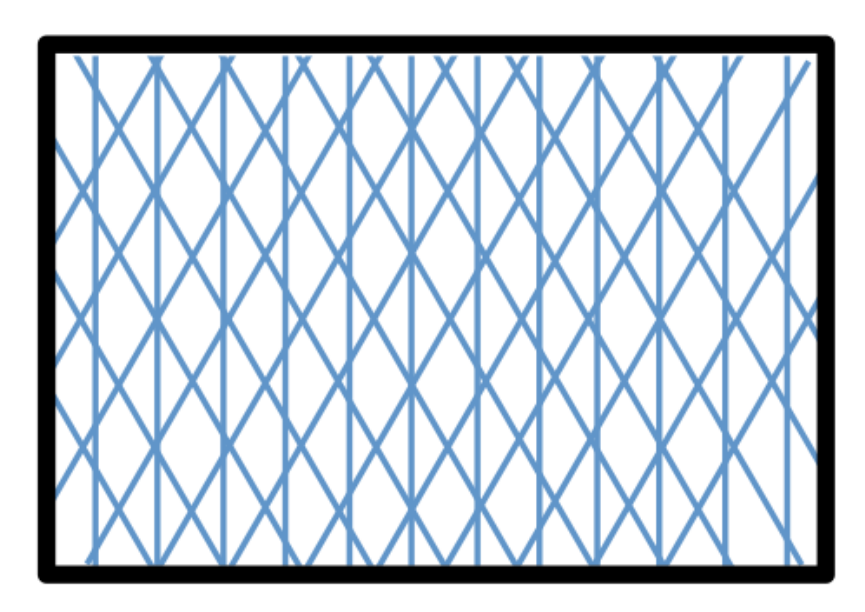}
%}}$
\caption{Left: Schematic view of the elements of a LAr TPC in two dimensions. The blue line represents a track traveling at an angle $\phi$ relative to the electric field direction. Ionization electrons produced on the track follow the electric field lines (yellow) through the induction planes until they are collected on wires in the collection plane. Right: View of the wire planes as seen by the approaching electrons illustrating that the electron trajectories shown in the left figure are overly simplistic.}
\label{fig:tpc}
\end{figure}

Electrons drift to the anode following the electric field lines as shown schematically in Figure \ref{fig:tpc} with a velocity of $\sim$1 mm/$\mu s$. Figure \ref{fig:driftvelocity} shows the electric field dependence of the electron drift velocity. A fraction are lost during transport to the anode wires due to  attachment on electro-negative impurities such as water and oxygen \cite{mts}. The charge loss is characterized by the drift electron lifetime, $\tau_e$. The ionization charge remaining after recombination, $Q_o$, is obtained from the collected charge at the wire planes, $Q_c$, using the relation $Q_o = Q_c / \exp(-(t_{arrive} - t_o) / \tau_e)$ where $t_{arrive}$ is the time of arrival of the electrons at the anode and $t_o$ is the time of ionization event. The value of $t_o$ may found from a beam timing signal as was done for ArgoNeut, from a photon detection system as is done in MicroBooNE or by selecting cosmic ray muons that pass through both the anode and cathode planes. Diffusion will increase the spatial extent of the electron cloud governed by the equation $\sigma_D = 2 \sqrt{D_{L(T)} (t_{arrive} - t_o) }$ where $D_{L(T)}$ is the longitudinal (transverse) diffusion coefficient \cite{shibamura}\cite{derenzo}\cite{cennini}\cite{atrazhev}. The  diffusion rms in MicroBooNE is expected to be 1.5 mm along the drift direction and 2.3 mm transverse to the drift direction.

The anode consists of several wire planes with bias voltages set so that ionization electrons travel between the wires of the first set of induction planes inducing a bipolar signal. The charge is collected on the last plane - the collection plane. The wire plane bias voltage settings for transporting 100\% of the ionization electrons to the collection plane is a function of the wire diameter, wire spacing and plane spacing \cite{buneman}.  Full transparency, or 100\% transmission of electrons to the collection plane, is achieved when the electric field between wire planes increases by $\gtrsim$ 40\% for each successive plane gap. There is a concomitant increase in the electron velocity of $\sim$15\% in each gap resulting in somewhat narrower wire signals in subsequent gaps. The wires in each plane are oriented by some angle relative to each other to provide a different view of the ionization event in each plane. The electrons will have a $\sim$50\% longer travel distance in 3D through the wire planes than indicated by the 2D representation shown in Figure \ref{fig:tpc}. 

Signals formed on anode wires are dominated by the motion of ionization electrons between the wire planes occurring on the time scale of microseconds. The positive leading lobe of the signal induced on the first instrumented induction plane, the U plane shown in Figure \ref{fig:tpc}, would be negligible if the Grid plane did not exist because the induced current would occur during the millisecond-scale drift time of electrons in the meter-scale main volume.  The negative trailing lobe is sizable since it is created from electrons traveling in the few mm gap between the U plane and the next (V) plane.  %These effects are illustrated in the MicroBooNE detector as shown in  Figure \ref{fig:rawsignals} which shows a wire signal from a through-going muon traveling almost parallel to the anode wires ($\phi \approx 90^\circ$).

For this report, we will assume that there are three planes, U (first induction), V (second induction) and W (collection). In the following, the ``field response'' refers to the time-dependent induced and collected charge on the wires due to the passage of ionization electrons through the wire planes.

The velocity of positive ions is significantly slower than that of electrons, approximately 5 mm/s, resulting in a positive ion buildup, or ``space charge'', that can affect the function of long drift TPCs operated in a high rate of background ionization. The electrical circuit is complete when ions reach the cathode plane. This may be a few minutes in a 2 meter drift TPC.

The wire plane configurations for several Fermilab detectors are shown in Table \ref{tab:config}, where Grid denotes an un-instrumented induction plane. A wire with $0^\circ$ orientation is vertical. Use of a grid plane restores the positive leading lobe effectively doubling the size of signals on the first instrumented induction plane at the expense of increasing the operating voltage of the wires.  %An additional benefit is that it protects against damage to readout electronics from electrostatic discharge during handling.

\begin{table}[tbp]
\caption{Wire Configuration.}
\label{tab:config}
\smallskip
\centering
\begin{tabular}{|l|l|}
\hline
Detector & Wire planes and orientation \\
\hline
ArgoNeuT & Grid(0$^\circ$), Induction(30$^\circ$), Collection(-30$^\circ$)\\
MicroBooNE & Induction(60$^\circ$), Induction(-60$^\circ$), Collection(0$^\circ$)\\
(Proto) DUNE & Grid(0$^\circ$), Induction(44$^\circ$), Induction(-44$^\circ$), Collection(0$^\circ$)\\
\hline
\end{tabular}
\end{table}

The ICARUS detector was instrumented with different charge integrating amplifiers on the middle induction plane to produce unipolar signals on all wire plane. Hits were then reconstructed by fitting to the function in equation \ref{eq:icarhit} to the raw wire signals.  

\begin{equation}
\label{eq:icarhit}
f(t) = B + A \frac{\mathrm{e}^{-\delta t / \tau_1}}{1 + \mathrm{e}^{-\delta t / \tau_2}} \quad \mathrm{where} \quad \delta t = t_{arrive} - t_o
\end{equation}

In contrast, the ArgoNeuT and MicroBooNE experiments elected to instrument all planes with the same electronics. As a consequence, some offline signal processing is desirable for reconstructing hits in the induction planes in these detectors. We will use the term ``electronics response'' to refer to the response of the full readout electronics chain by the injection of a $\delta$-function input.

%\begin{figure}[tbp] 
%\centering
%\includegraphics[width=.9\textwidth]{uBWireSignals.pdf}
%\caption{Pedestal-subtracted raw signals from a cosmic ray muon on a wire in the first induction plane (U, top),  a wire in the second induction plane (V, middle) and a wire in the collection plane (W, bottom) in the MicroBooNE detector. The horizontal axes are the sample number in the TPC readout window, or ``tick'', where 1 tick = 0.5 $\mu s$. The vertical axes are the pedestal subtracted digitized signals (ADC counts). The signals are not from the same ionization event as can be inferred by noting the large difference in the arrival time.}
%\label{fig:rawsignals}
%\end{figure}

\begin{figure}[tbp] 
\centering
\includegraphics[width=\textwidth]{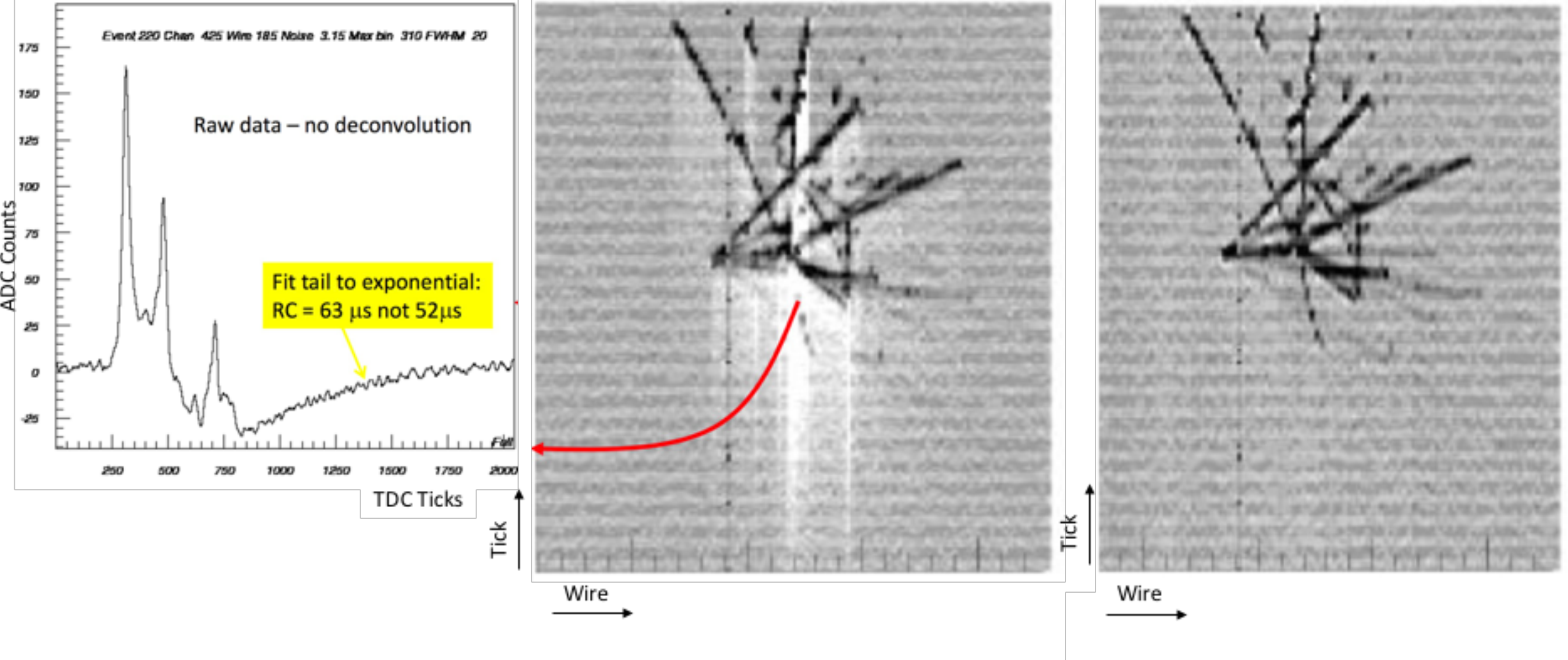}
\caption{Neutrino interaction with secondary hadronic interaction in the ArgoNeuT detector. Data were taken in the NuMI beam. Left: Raw ADC counts vs time tick (1 tick = 0.2 $\mu$s) for a wire in a high ionization region in the collection plane indicated by the red arrow. The RC time constant of the tail was found to be 63 $\mu$s instead of 52 $\mu$s from bench measurement presumably due to additional capacitance of the in-situ cables. Middle: Time vs wire number (wire spacing = 4 mm) of raw ADC data. The image grey scale is the raw ADC values. The white regions show a baseline shift due to the mis-match of the readout electronics. Right: The same event after deconvolution.}
\label{fig:argodisplay}
\end{figure}

Raw signals in the ArgoNeuT detector in Figure \ref{fig:argodisplay} show the effects of an impedance mismatch between the TPC wires and the readout electronics. The readout electronics were borrowed from another operating experiment with the condition that they would not be altered. Bench measurements showed a significant baseline undershoot with a time constant of 52 $\mu$s. In-situ measurement of the time constant was found to be 63 $\mu$s. The right hand image in this figure shows the same event after deconvolution using the method described in the next section.

More precise values of the order of magnitude estimates used in this section are calculated in an Excel spreadsheet that includes parameterizations of liquid argon properties such as those shown in Figure \ref{fig:driftvelocity}.  This ``LAr TPC Calculator'' is described in Appendix \ref{sec:calc}. 

\section{Offline Signal Processing}\label{sec:osp}

Detector features and electronics artifacts described in the previous section can be mitigated by offline signal processing. The digitized signal is a convolution of the serial effects of signal formation, electron transport through the TPC and processing by the readout chain as shown schematically here: Wire Signal = Ionization $\bigotimes$ Recombination $\bigotimes$ Diffusion and Attachment $\bigotimes$ Field Response $\bigotimes$ Noise $\bigotimes$ Electronics Response.  In this section we describe a Fourier deconvolution method that removes the most deleterious effects to prepare signals for hit finding. This method also provides a mechanism for removing coherent noise using a frequency-domain Wiener filter \cite{wiener}. 

We can estimate the frequency range of wire signals by using a Gaussian approximation where the width of the signal in the frequency domain, $\sigma_f = 1 / (2 \pi \sigma_t)$, where $\sigma_t$ is the width in the time domain. The signal in the frequency domain is also a Gaussian having a maximum at $f = 0$.  An estimate of the high frequency cut-off can be made by observing that the highest frequency component of a wire signal is related to the inverse of the transit time through the wire plane gap. Using an example where $\sigma_t$ = 2 $\mu$s, a low-pass Wiener filter should have a 3$\sigma_f$ cut-off at $\approx$200 kHz.
 
It is important that the filter preserves the most desirable components of the wire signal. The filter may be adjusted to remove lower frequency components resulting in narrow signals in the time domain in an effort to improve the separation of close tracks. This may have a detrimental effect on calorimetric reconstruction of the particle energy however since some of the signal power will be lost in the process. 

\subsection{Electronics Response} 

\begin{figure}[tbp] 
\centering
\includegraphics[width=.8\textwidth]{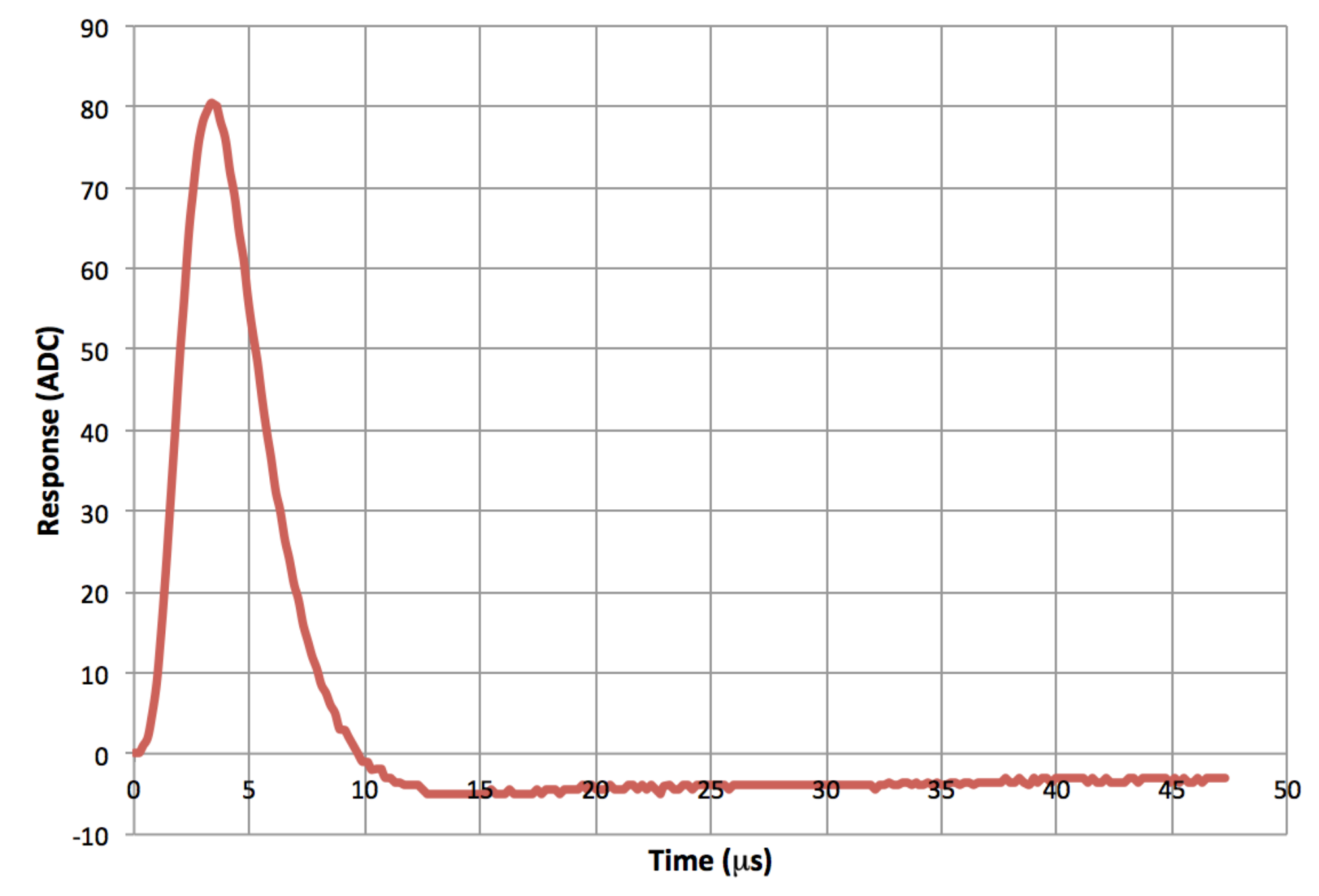}
\caption{Response of the ArgoNeuT preamplifier to a $\delta$-function input. The scatter in the points is due to  digitization error of the ADC. }
\label{fig:argopreamp}
\end{figure}

The electronics response of a $\delta$-function input is generally obtained from simulations and bench tests. Figure \ref{fig:argopreamp} shows the output of the bench test of the ArgoNeuT preamplifier when a $\delta$-function signal is injected. The data were taken with the ArgoNeuT DAQ system consisting of 10 bit ADCs sampled at 0.2 $\mu$s/tick. The noise rms was $\approx$1 ADC count as can be inferred by the jitter on the long tail. The large baseline shift  near 15 $\mu$s  in Figure \ref{fig:argodisplay} is due to the impedance mis-match. 

\begin{figure}[tbp] 
\centering
\includegraphics[width=.8\textwidth]{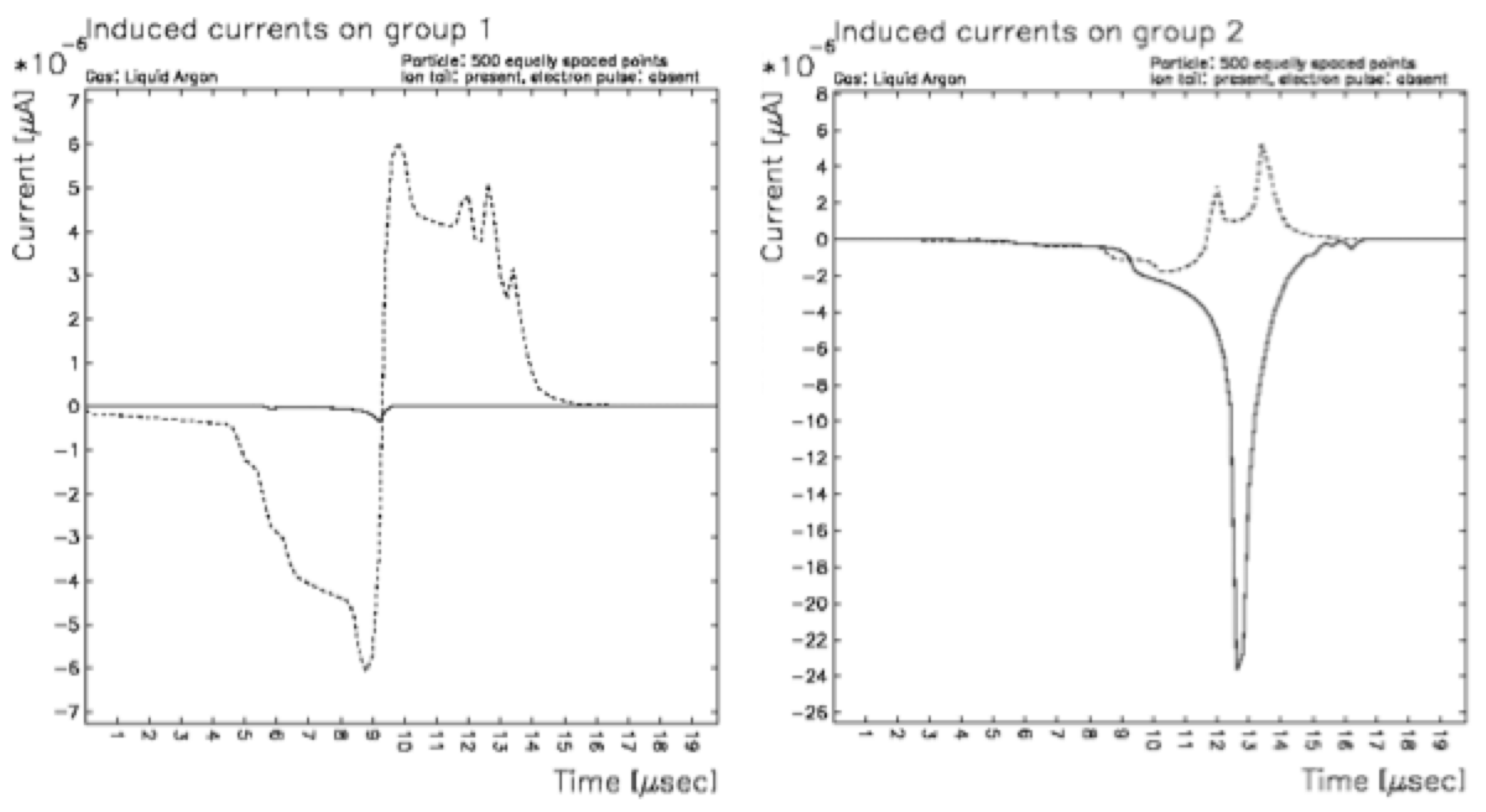}
\includegraphics[width=.8\textwidth]{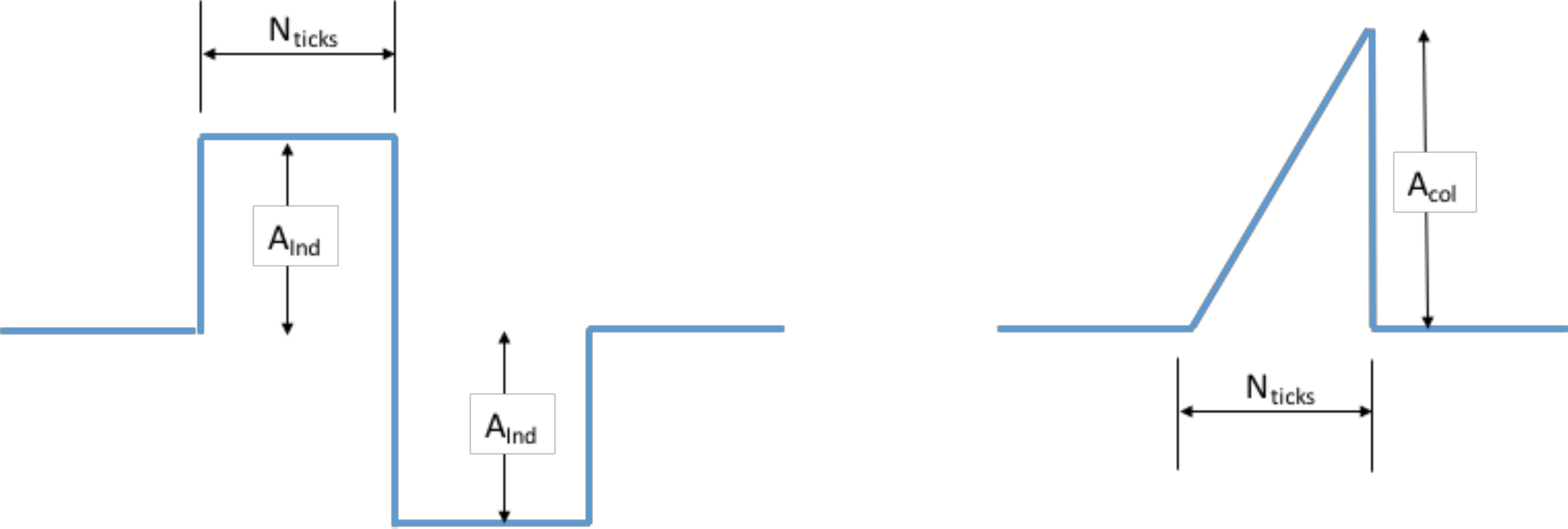}
\includegraphics[width=.8\textwidth]{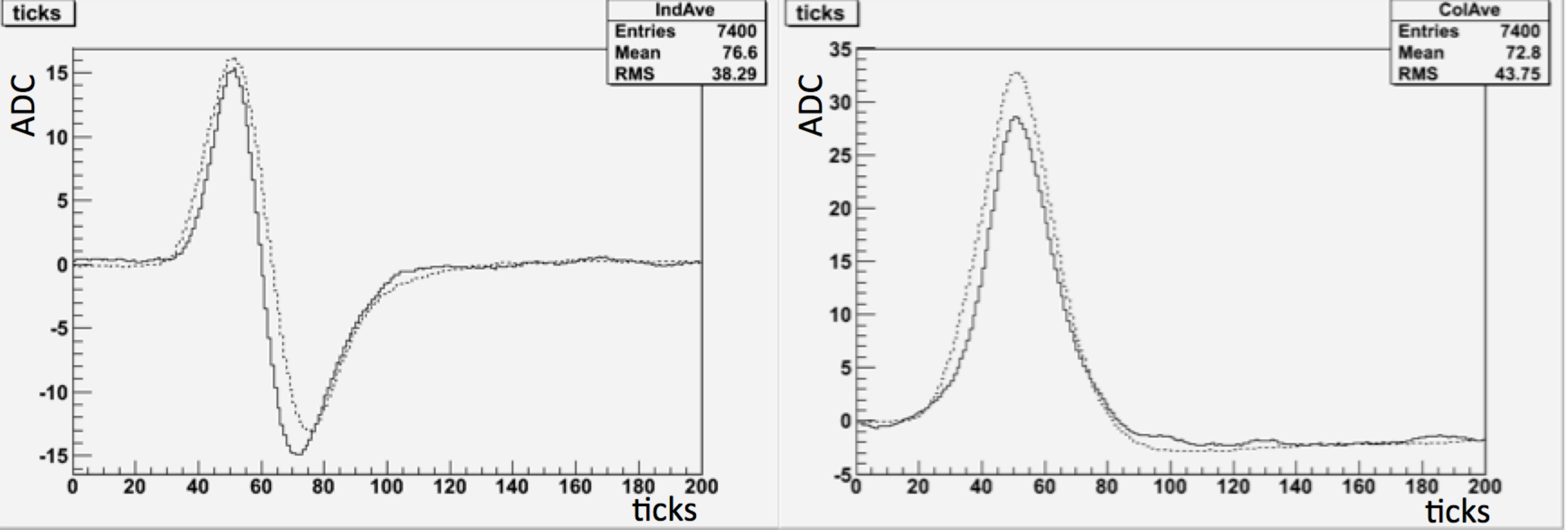}
\caption{Top: Garfield simulation of the current on the induction plane in ArgoNeuT (Left) and in the collection plane (Right) for a track with $\phi$ = 90$^\circ$. Solid (dotted) black lines represent the direct (induced) signal. Middle: Simplified representation of the Garfield field response used in the Monte Carlo simulation and deconvolution. The values of N$_{ticks}$, A$_{ind}$ and A$_{col}$ were determined by matching simulated and real data. Bottom: Real detector wire signals (solid) and simulated wire signals (dashed).}
\label{fig:garfield}
\end{figure}

\subsection{Field Response} 

Determining the field response appears to be a difficult problem. Ionization electrons follow complex 3D trajectories as they pass through the wire planes producing direct and induced signals on nearby wires. Signals are induced on neighboring wires due to inter-wire capacitance. An additional complication is that the detailed shape of the field response depends on the wire bias voltage settings. 

In practice it is generally sufficient to model the field response with a simple analytic form that can be scaled for different  operating conditions. The method used in ArgoNeuT is illustrated in Figure \ref{fig:garfield}. The broad features of wire signals simulated by Garfield \cite{garfield} shown in the top panels are represented by the simple forms shown in the middle panel that are scaled in time and amplitude to match wire signals in the real detector (lower panel). The real detector data shown in the lower panel (solid lines) were obtained by averaging wire signals on $\approx$50 wires on a through-going muon with $\phi \approx 90^\circ$. The real muon was reconstructed in 3D and then simulated with the ArgoNeuT Monte Carlo. The dashed lines are the average of 50 wire signals on the simulated muon where the induction (collection) plane, $N_{ticks}$ was set to the 1.5 (2.5) $\times$ the width obtained from the 2D Garfield simulation. The field response amplitudes, $A_{Ind}$ and $A_{Col}$, were then adjusted and an asymmetry applied to the induction plane response to improve the agreement. The use of such a simple model is only possible when the time constant of the electronics response exceeds that of the field response.

\subsection{Diffusion and Attachment} 

The effects of diffusion are generally small. Using MicroBooNE as an example, the longitudinal diffusion rms for electrons traveling the full 2.5 m drift is $\sim$1.5 mm corresponding to an increase in the time spread of 1 tick. Likewise, correcting for electron attachment is a simple multiplicative factor that can be applied in later stages of reconstruction.

\subsection{Recombination} 

A recombination correction is required for calorimetric reconstruction but that requires knowledge of the path length of the track in space to determine $dQ_o/dx$. It is important to recognize that this correction is unavoidably imprecise since recombination occurs on the physical length scale of microns in the TPC while the correction is applied to the collected charge $Q_o$ with the length scale of the wire spacing. The disparity in the length scales is not important when $dE/dx$ is slowly varying but is significant near the Bragg peak. A related imprecision occurs in a Monte Carlo simulation of the charge deposition. As an example, Geant4 \cite{geant} provides the energy lost by a particle in a tracking step of size $\sim$1 mm. Applying the recombination correction results in a value of $dQ_o/dx$ that depends on the tracking step size and the variation of $dE/dx$ within the step.

Applying a recombination correction to hits reconstructed in electromagnetic showers requires a different but ultimately simpler treatment. The example of a low energy photon conversion where the e$^+$e$^-$ pair have an opening angle of $\sim$0.01 radians illustrates this point. The pair initially do not have a dipole field and hence no ionization occurs. After traveling $\mathcal{O}$(10) nm, the dipole separation is about 1 atomic diameter resulting in the onset of two overlapping ionization columns in which recombination occurs between electrons and ions produced by both particles. This situation is somewhat similar to that of a single particle with twice the $dE/dx$ of a minimum ionizing particle. After traveling $\mathcal{O}$(100) $\mu$m, the ionization columns are sufficiently well separated such that recombination occurs preferentially between electrons and ions produced in the same column. For this pair-production example, the charge measured on a wire is largely due to two well-separated minimum ionizing particles. As a result, it is reasonable to use a recombination factor of $\mathcal{R}$ = 0.64 for minimum ionizing particle hits that reside in the shower. This approximation clearly fails if the local charge density is very high.

\subsection{Signal Deconvolution}

\begin{figure}[tbp] 
\centering
\includegraphics[width=0.8\textwidth, angle=-90]{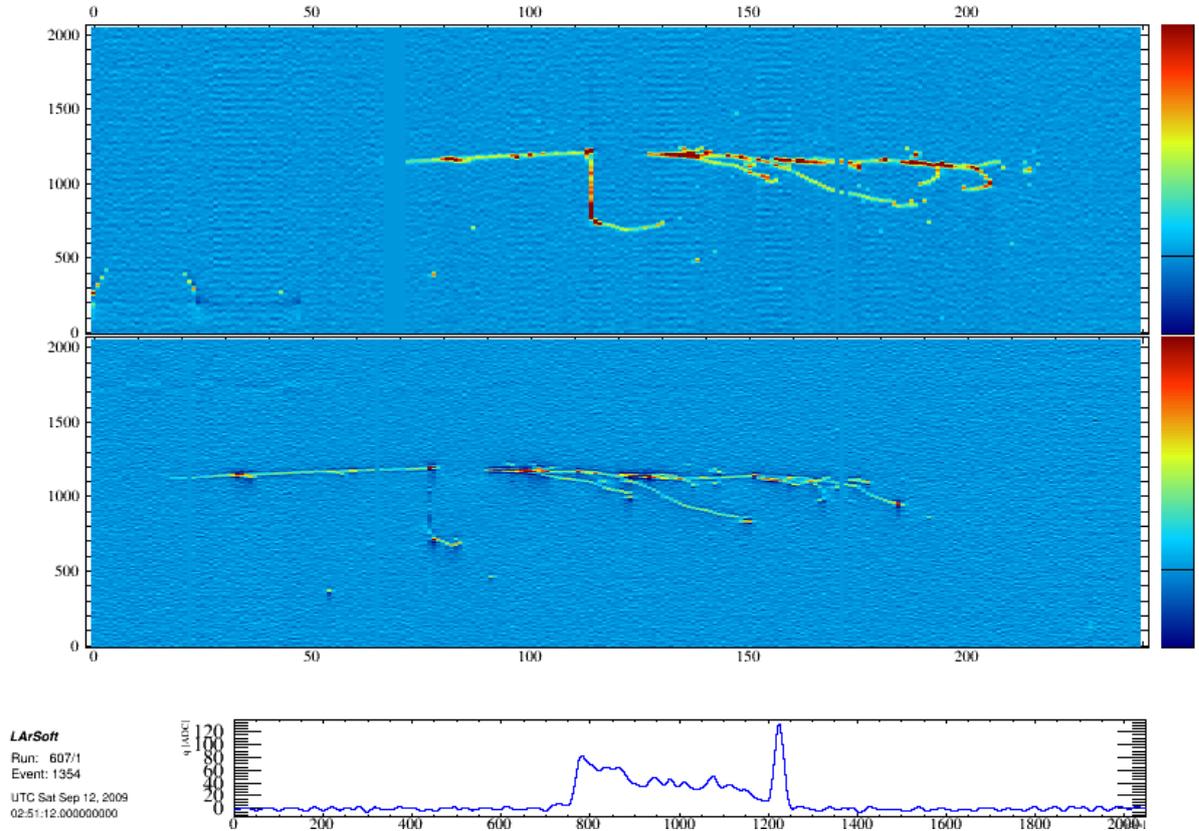}
\caption{Top: Collection plane view of an interaction in the ArgoNeuT detector producing a particle that travels along the electric field direction towards the anode plane. The particle travels for 10 cm before it decays or re-interacts. Bottom: ADC vs time of the collection plane wire along which the particle travels. The large peak near 1200 ticks is due to low energy particles produced in the collision. Ionization fluctuations along the particle trajectory are apparent as it travels from 1200 $\rightsquigarrow$ 800 ticks. Middle: Induction plane view showing the (expected) negligible signal. The wire signals are shown after deconvolution.}
\label{fig:smallphi}
\end{figure}

The approach described here is to remove the electronics response and field response, resulting in wire signals that are roughly Gaussian in shape in all wire planes. A fit to a Gaussian distribution provides a good estimate of the hit position and the ionization charge. Ionization electrons produced on tracks that travel roughly in  the electric field direction, $\phi \approx 0$, will arrive over many microseconds. In this case a single Gaussian fit is clearly inadequate. An extreme example from the ArgoNeuT data is shown in Figure \ref{fig:smallphi} in which a track with $\phi = 0$ is evident on wire 113 in the collection plane. A Gaussian fit would not provide the information required to reconstruct this track.

The time sampled wire signal read out by the data acquisition system, denoted $W(t)$, is the convolution of the ionization charge approaching the wire plane, $Q(t)$, with the field response, $F(t)$, and the electronics response $E(t)$. 

\begin{equation}
\label{eq:convolution}
W(t) = Q(t) * F(t) * E(t)
\end{equation}

\noindent  The time-dependent ionization charge can be recovered by deconvolution

\begin{equation}
\label{eq:deconvolution}
Q(t) = \mathpzc F^{-1}  \Bigg[ \frac{\Phi(f) \times {\mathpzc F}(W(t))}{ {\mathpzc F}(F) \times {\mathpzc F}(E)} \Bigg]
\end{equation}

\noindent where $\mathpzc F$ is the Fourier Transform and $\Phi (f)$ is a Weiner filter function. Here we have dropped the $t$ dependence of $F$ and $E$ to distinguish the real-time varying wire signal from the real-time invariant response functions. A deconvolution kernel that includes the filter, field response and electronics response is computed for each wire plane.

The ArgoNeuT data acquisition system acquired 2048 samples at 5MHz for each NuMI beam spill. A standard Fast Fourier Transform (FFT) requires 2$^N$ samples so no zero-padding was required. The electronics response from Figure \ref{fig:argopreamp}, the field response from Figure \ref{fig:garfield} and the filter function were sampled or calculated at the same sampling rate and zero-padded to produce the deconvolution kernel of size 2048. Each of the 480 wires in the ArgoNeuT detector were then processed using Equation \ref{eq:deconvolution}. 

Signal deconvolution is now a standard feature of the wire calibration module in LArSoft. This procedure was found to be computationally expensive for MicroBooNE however. The MicroBooNE data acquisition system acquires 9600 samples at 2 MHz. Zero-padding to 2$^N$ samples for the FFT requires complex arrays of size 16384. A significant performance improvement is gained by processing only those regions which have sizable signals.  We define a ``Region of Interest'', or ROI, which is a block of consecutive ADC samples which are above (below) a positive (negative) noise threshold plus pre-padding and post-padding with some number of ADC samples. Multiple ROIs were packed into fixed-size arrays of size 2048 and deconvolved. A pedestal subtraction in each ROI was then done to remove DC offsets. %Figure \ref{fig:rawdeconroi} shows an example of a simulated MicroBooNE V-plane raw signal and the result after application of this procedure. The wiggles in the baseline are due to a non-optimum filter which did not adequately remove noise. Work by others in the MicroBooNE collaboration have recently improved the filter function that greatly reduces this effect. 
%The ROI algorithm can reduce the amount of stored data by a factor of $\approx100$.

The Wiener filter function is constructed to allow for ROIs of variable length. The Wiener filter has the form

\begin{equation}
\label{eq:filter}
\Phi(f) = \Bigg[  \frac{S^2(f)} {S^2(f) + N^2(f)}  \Bigg]
\end{equation}

\noindent where $S(f)$ is the Power Spectral Density (PSD) of the Signal and $N(f)$ is the PSD of the noise. The filter is constructed by creating histograms of PSD for ROIs that contain the wire signal, $S(f)$, of a small angle track and a sideband histogram that has no detectable signal, $N(f)$. The results of equation \ref{eq:filter} are then cast into a function that allows scaling $\Phi(f)$ for ROIs of different length.

%\begin{figure}[tbp] 
%\centering
%\includegraphics[width=0.9\textwidth]{RawDecon.pdf}
%\caption{Top: Simulated raw signal in the V plane of a shallow angle track in the MicroBooNE detector. A Region Above Threshold (RAT) is found where the absolute value of the wire signal is $ 0 \pm 5$ ADC. The RAT is padded at both ends to create a Region of Interest (ROI). Bottom: ROI after deconvolution.}
%\label{fig:rawdeconroi}
%\end{figure}

\section{Hit Reconstruction}\label{sec:hit}

Hit reconstruction is the seemingly straightforward process of characterizing the position and amount of charge deposited in the detector within a ROI. The hit charge is simply the integral of a wire signal ROI times a calibration constant. The hit time relative to $t_o$ could be calculated using the charge weighted mean of the ROI. Information that could be used to separate close tracks or complicated ionization events would be lost using this simple procedure however. The method described here improves the reconstruction of hits; in particular those close to a neutrino interaction.

Each ROI is fitted to a variable number of Gaussian distributions defined by a time, peak  amplitude and width. The first step is to estimate the values of these parameters. A set of local peaks is found in the ROI. Each peak is then fitted to a Gaussian distribution with that starting value. An initial fit is performed assuming the noise rms is 1 ADC count. A cut on the $\chi^2$/DOF of the fit compensates for a non-Gaussian field response and for the noise rms $\ne$ 1. The $\chi^2$/DOF of the fit is used to decide whether to continue fitting with additional ``hidden'' Gaussian distributions that do not have a local maximum or to create a ``crude hit'' that encapsulates the global features of an ionization event in this ROI. %The signal in the bottom panel of figure \ref{fig:smallphi} clearly falls in the latter category.

\begin{figure}[H] 
\centering
\includegraphics[width=0.9\textwidth]{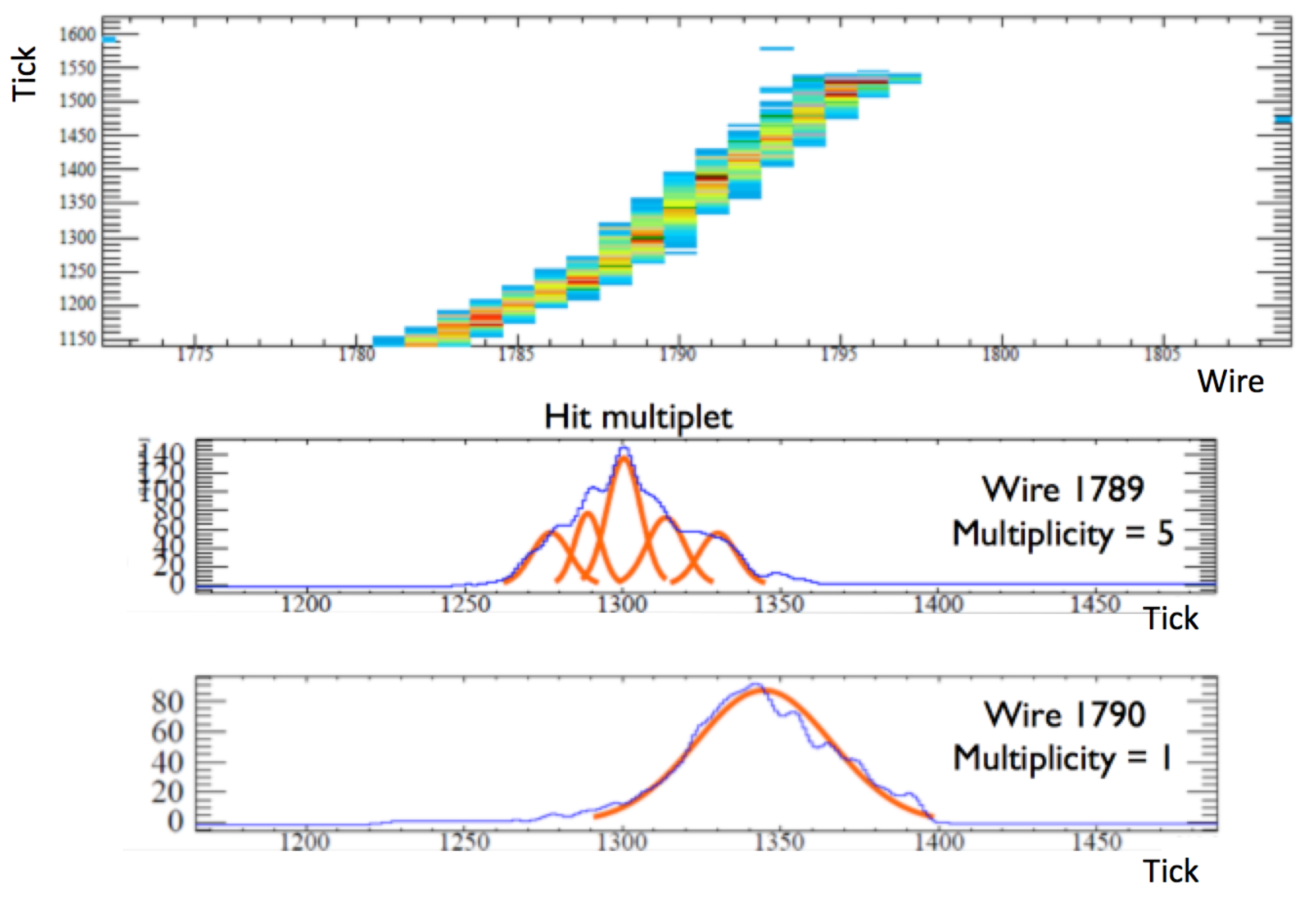}

\caption{Top: A cluster of hits from a simulated large angle stopping track. Middle: A multiplet of 5 hits (orange curves) created on wire 1789. Bottom: A single hit created on the next wire.}
\label{fig:hitmultiplet}
\end{figure}

The properties of a hit are the parameters from the Gaussian fit. A hit is considered to be a member of a ``multiplet'' if it was found in a multi-Gaussian fit. Figure \ref{fig:hitmultiplet} illustrates the rationale for introducing this concept. The top panel shows the trajectory of a simulated low momentum particle in the collection plane. Gaussian fits to the charge deposited on two adjacent wires are shown in the middle and lower panels. A good fit to five Gaussian distributions was found on one wire which resulted in a multiplet of five hits. A good fit to one Gaussian distribution was found on the adjacent wire resulting in one hit.  

It is clear that there is insufficient information at this stage of reconstruction to make an unambiguous definition of a hit that represents the deposited charge. The 5-hit multiplet in this figure could conceivably have been created by multiple tracks created in a neutrino interaction or some other complex interaction. 

A second type of ambiguity arises when charge from several tracks overlap and are fit as a single Gaussian hit. This occurs at the primary vertex of every neutrino interaction to some extent. One negative consequence is that pattern recognition of tracks near the vertex will be incorrect. Another  consequence is that the calculation of $dQ/dx$ of short tracks near the primary vertex will be erroneous. Untangling the effects of overlapping tracks should potentially allow the identification of MeV-scale particles produced by final state interactions.

\section{Summary}\label{sec:summary}

The offline signal processing techniques described in this report are a significant first step on the path to a  fully automatic reconstruction of complex interactions in a LAr TPC. %Additional development is clearly required, particularly in the areas of hit definition and better processing of induction plane wire signals. The motivation is clear for pursuing an iterative approach to refine information obtained during the early stage of reconstruction using information obtained in later stages.

%\pagebreak
\appendix
\section{Appendix A - LAr TPC Calculator}\label{sec:calc}

The LAr TPC calculator is an Excel spreadsheet that is useful for understanding the operation of a LAr TPC. Basic detector parameters such as the wire spacing, wire diameter, drift electric field, etc, are entered resulting in the calculation of measured detector quantities. For example, the wire signal amplitude on the collection plane produced by an idealized track that travels perpendicular to the collection plane wires is calculated. This is the smallest signal of interest for detectors used in neutrino experiments. The calculator utilizes a parameterization of the MicroBooNE and LArIAT cold ASIC preamplifier that was developed at Brookhaven National Laboratory \cite{asic}, a second stage amplifier and a generic sampling ADC. 

Derived quantities such as the drift electron velocity, maximum drift time and signal to noise ratio are displayed in blue shaded cells. Non-scaling properties of liquid argon such as the ionization energy and diffusion coefficients are highlighted by the salmon colored cells.

The calculator accounts for the electric field dependence of the electron drift velocity. The charge loss due to recombination is calculated using both Birk's Law and Modified Box Model formulas. The plots shown in figure \ref{fig:driftvelocity} were produced by the calculator.

\begin{figure}[h] 
\centering
\includegraphics[width=1.15\textwidth]{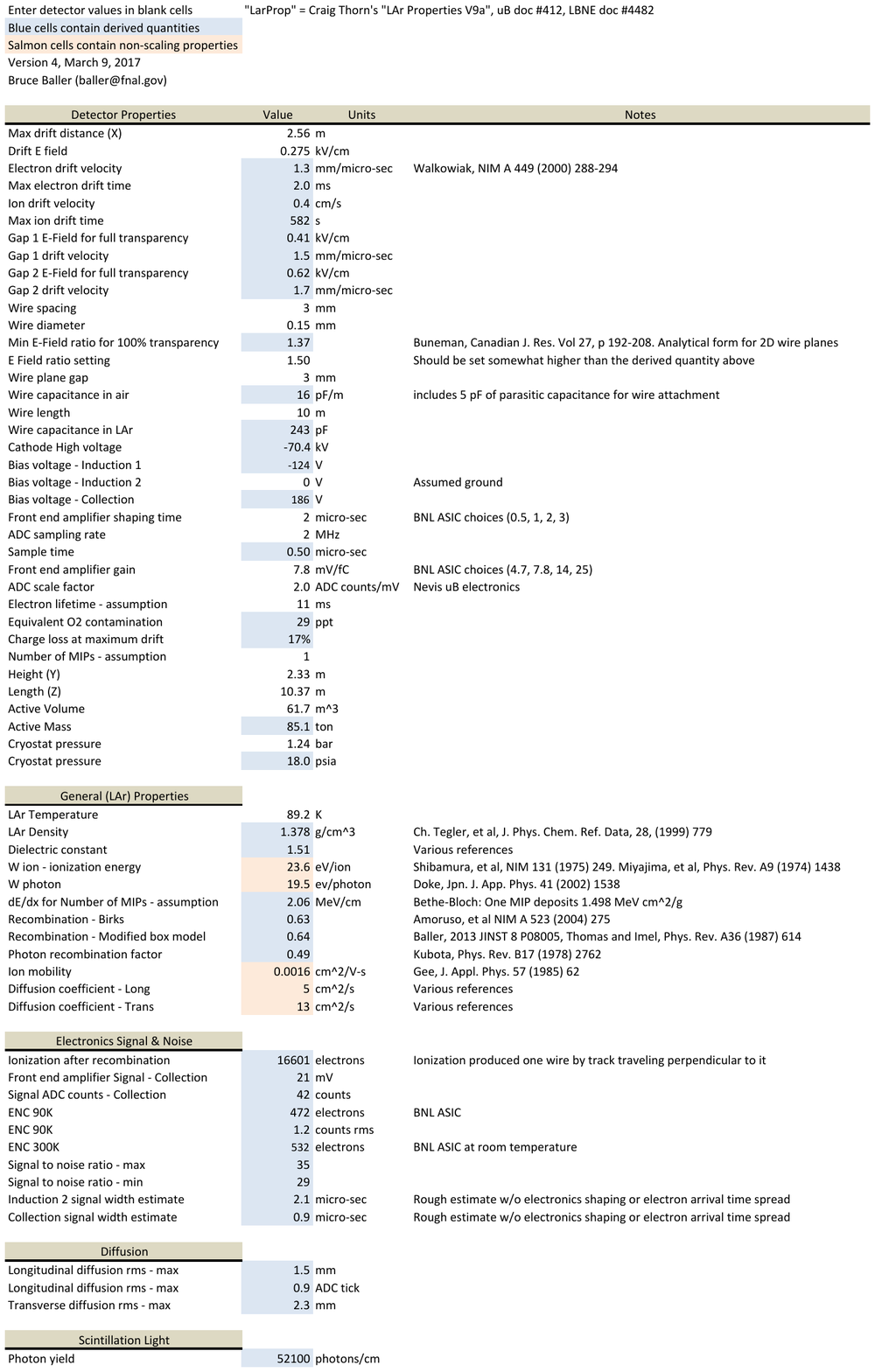}
\caption{LAr TPC calculator configured with MicroBooNE detector parameters.}
\label{fig:calc}
\end{figure}

\acknowledgments

Members of the ArgoNeuT and MicroBooNE collaborations provided many helpful comments during the development of these techniques and continue to make further improvements. Carl Bromberg and Dan Edmunds from Michigan State University provided insightful guidance during the development of the deconvolution method.

We gratefully acknowledge the support of Fermilab, the U.S. Department of Energy and the National Science foundation. Fermilab is operated by Fermi Research Alliance, LLC under Contract No. DE-AC02-07CH11359 with the United States Department of Energy.

\end{document}